\documentclass{IEEEtran4PSCC}
\usepackage{transparent}

\usepackage{balance}
\usepackage[hyphens]{url}
\usepackage{hyperref}

\ifCLASSINFOpdf
   \usepackage[pdftex]{graphicx}
   \usepackage{grffile}
   \usepackage{import}
  \graphicspath{{figs/}} 
\else
   \usepackage[dvips]{graphicx}
\fi
\usepackage{subcaption} 

\usepackage[cmex10]{amsmath}
\usepackage{tikz,xcolor,hyperref}

\makeatletter
\let\old@ps@headings\ps@headings
\let\old@ps@IEEEtitlepagestyle\ps@IEEEtitlepagestyle
\def\psccfooter#1{%
    \def\ps@headings{%
        \old@ps@headings%
        \def\@oddfoot{\strut\hfill#1\hfill\strut}%
        \def\@evenfoot{\strut\hfill#1\hfill\strut}%
    }%
    \def\ps@IEEEtitlepagestyle{%
        \old@ps@IEEEtitlepagestyle%
        \def\@oddfoot{\strut\hfill#1\hfill\strut}%
        \def\@evenfoot{\strut\hfill#1\hfill\strut}%
    }%
    \ps@headings%
}
\makeatother

\psccfooter{%
        \parbox{\textwidth}{\hrulefill \\ \small{Preprint}}%
}

\begin{document}
%
\title{Stability Assessment Considering Multi-Timescale Power System Dynamics: Insights into Hopf Bifurcations with GFL and GFM IBRs}


\author{
\IEEEauthorblockN{Luis David Pabón Ospina}
\IEEEauthorblockA{Grid Control and Grid Dynamics\\
Fraunhofer IEE\\
Kassel, Germany\\
luis.david.pabon.ospina@iee.fraunhofer.de}
\and
\IEEEauthorblockN{Martin Braun}
\IEEEauthorblockA{Fraunhofer IEE\\
University of Kassel\\
Kassel, Germany\\
martin.braun@uni-kassel.de}
\and
\IEEEauthorblockN{Sushobhan Chatterjee\\Sijia Geng}
\IEEEauthorblockA{Dept. of Electrical and Computer Engineering\\
Johns Hopkins University\\
Baltimore, MD, USA\\
\{schatt21, sgeng\}@jhu.edu}
}


\maketitle

\begin{abstract}
Real power systems exhibit dynamics that evolve across a wide range of timescales, from very fast to very slow phenomena. Historically, incorporating these wide-ranging dynamics into a single model has been impractical. As a result, power engineers rely on timescale decomposition to simplify models. When evaluating fast phenomena, slow dynamics are often neglected (assumed stable), and vice versa. This paper challenges the prevailing paradigm by demonstrating the importance of assessing power system stability while considering multiple timescales simultaneously. Using the concept of Hopf bifurcations, it exemplifies instability issues that would be missed if multi-timescale dynamics are not considered. Although this work employs both grid-following and grid-forming inverter-based resource models, it is not a direct comparison. Instead, it presents a case study demonstrating how one technology can complement the other from a multi-timescale dynamics perspective.
\end{abstract}


\thanksto{\noindent This work was supported in part by the German Ministry for Economic Affairs and Energy, in part by the Projekträger Jülich under Projects LISA -FKZ03EI4059A and SysStab2023 -FKZ03EI6122H. The JHU team was supported by the Johns Hopkins Ralph O'Connor Sustainable Energy Institute (ROSEI). Only the authors are responsible for the content of this publication.}

\section{Introduction}
The adoption of inverter-based resources (IBRs) has introduced significant complexities into power system analysis, control, and operation \cite{geng2025unified}.
Even in traditional power systems without IBRs, it was impractical to incorporate dynamics from a wide range of timescales into a single model. For example, when assessing rotor angle stability, it is common practice to consider only the dynamics in the timescale of electromechanical oscillations. Models for automatic voltage regulators (AVR), speed governors, and power system stabilizers (PSS) were typically considered in these studies, with simulations usually running for ten seconds or less. Slower dynamics, such as load restoration processes via automatic load tap changers (LTCs) and the action of over-excitation limiters (OELs), which act over tens of seconds or minutes, were usually neglected.

While this is a reasonable approach in some cases, timescale decomposition is often performed by system operators and utilities intuitively with little to no mathematical rigor. Nevertheless, power systems have interdependent dynamics across multiple timescales. This interdependency means that behavior from one timescale can cause issues in another. {A strong example of short-term instability induced by long-term dynamics is provided in Chapter 8.2.3 of \cite{V_Stab_Elec_PS}. In this case, the evolution of unstable slow variables leads to a short-term instability.} Three types of such instability are distinguished \cite{V_Stab_Elec_PS}:

\begin{itemize}
\setlength\itemsep{1em}
    \item S-LT1: The loss of short-term equilibrium caused by long-term dynamics. A typical example is the loss of synchronism of a synchronous generator after its OEL has limited the field current. This often results in an exponential collapse, similar to a saddle-node bifurcation or a purely real unstable eigenvalue. An example of this was illustrated in \cite{PabonOspina2025} by the authors.
    \item S-LT2: The loss of attraction to the stable short-term equilibrium due to a shrinking region of attraction caused by long-term dynamics.
    \item S-LT3: The oscillatory instability of short-term dynamics caused by long-term dynamics.
\end{itemize}

The last of these, S-LT3, is the main focus of this paper. In traditional systems, S-LT3 was plausible but rare. One possible scenario involved the OEL limiting the field current, which could inactivate the PSS that relies on the field current for its stabilizing action \cite{CIGRE1995}. With the growing integration of IBRs into power systems, it is important to revisit this mechanism to identify the risks associated with timescale simplifications and to prevent potential instabilities.

\section{Preliminaries on Nonlinear Dynamics}

In dynamical systems, equilibrium points can be created or destroyed, or their stability can change. These qualitative changes in the dynamics are called bifurcations, and the parameter values at which they occur are known as bifurcation points \cite{strogatz2015nonlinear}. In power system dynamics, two types of bifurcations are common: saddle-node bifurcation (SNB) and Hopf bifurcation. SNB occurs when a system's equilibria are suddenly created or destroyed as a parameter changes, and is often seen in power systems, such as the folding point of PV curves. From a stability perspective, this represents an eigenvalue of the state matrix crossing the imaginary axis at the origin (no imaginary part). Reference \cite{chatterjee2025voltage} shows the irrelevance of fast line dynamics for SNB in a broad class of IBRs. Hopf bifurcation happens when a stable fixed point loses its stability and gives rise to a periodic oscillation known as a \textit{limit cycle} \cite{seydel2009practical}. It can be either supercritical (stable limit cycle), or subcritical (unstable limit cycle), and is a mechanism by which a system can transition from a steady state to an oscillating state represented by a pair of complex conjugate eigenvalues of the state matrix crossing the imaginary axis. Reference \cite{chatterjee2024effects} explores the impacts of fast line dynamics for Hopf bifurcation in GFM IBRs. In this paper, we demonstrate that oscillatory instability resulting from a Hopf bifurcation induced by long-term dynamics can be a significant threat in modern power systems with IBRs. To explore this, we first introduce the concept of stable manifolds by considering the following dynamical system,
\begin{align}
    \dot{x} &= f(x,y), \label{eq:x} \\ 
    \epsilon \dot{y} &= g(x,y), \label{eq:y}
\end{align}
where \( x \) is the ``slow" state vector that contains all slow dynamics, \( y \) represents the ``fast" state vector, and \( \epsilon \) is a small parameter. The stable manifold is defined as the set of initial conditions $x_0$ such that $\mathbf{x}(t) \to \mathbf{x}^{\ast} \text{ as } t \to \infty$, where $\mathbf{x}^{\ast}$ is a fixed (or equilibrium) point.

\begin{figure}[!t]
    \centering
    \includegraphics[width=0.7\linewidth]{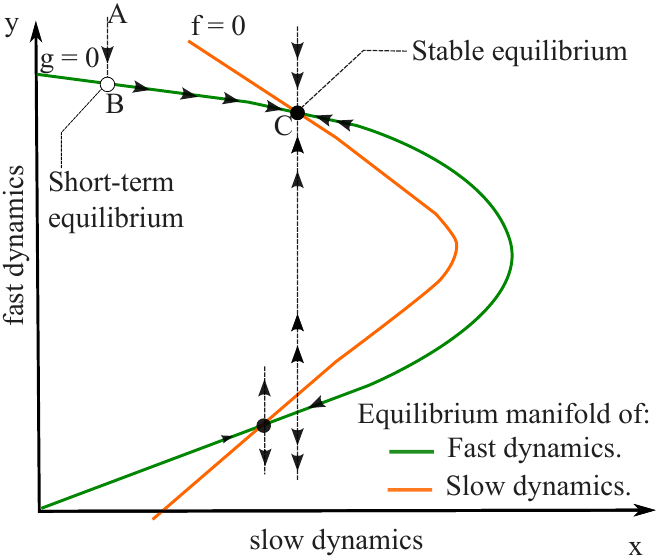}
    \caption{Equilibrium manifolds that intercept \cite{pes_tr9}.}
    \label{fig: manifolds_intercept}
\end{figure}

\begin{figure}[!t]
    \centering
    \includegraphics[width=0.7\linewidth]{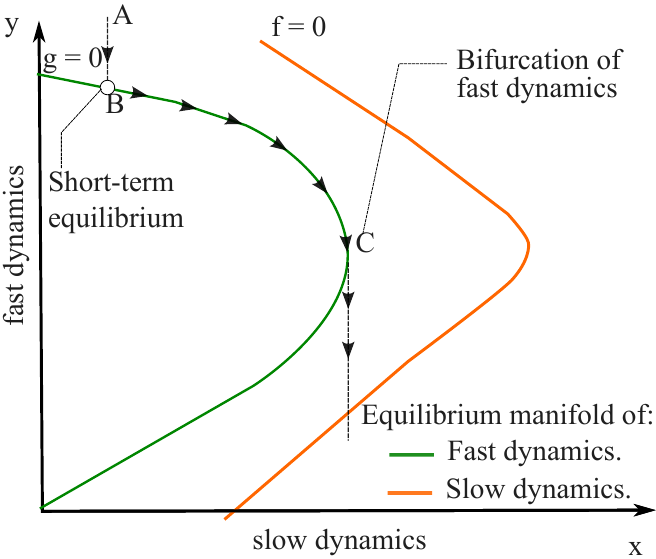}
    \vspace{0.4mm}
    \caption{Equilibrium manifolds that do not intercept \cite{pes_tr9}.}
    \label{fig: manifolds_dont_intercept}
\end{figure}

The equilibrium manifolds are illustrated in Figs.~\ref{fig: manifolds_intercept} and~\ref{fig: manifolds_dont_intercept}, as adapted from \cite{pes_tr9}.
Referring to Fig.~\ref{fig: manifolds_intercept}, consider that the system is operating at point A. If a disturbance occurs, such as a short circuit followed by a line trip, the system will rapidly transition from point A to point B, which lies on the fast dynamics equilibrium manifold (green curve), defined by \( g = 0 \) in \eqref{eq:y}. Point B represents a short-term equilibrium in which only fast dynamics are at a fixed point. After some time, the slow dynamics begin to influence the system, slowly drawing it towards the slow dynamics equilibrium manifold (orange curve). Eventually, the system reaches point C, where both \( f \) and \( g \) are zero in \eqref{eq:x} and \eqref{eq:y}. This represents the true equilibrium of the system \cite{PabonOspina2025}.

Nevertheless, both equilibrium manifolds don't always intercept. In this case, as shown in Fig.~\ref{fig: manifolds_dont_intercept}, the system will quickly transition from A to B, settling into a short-term equilibrium. Eventually, the slow dynamics act and push the system towards the slow dynamics equilibrium manifold. Before the system can settle into a point where $f = 0$, there is a bifurcation of the fast dynamics at point C and the system becomes unstable.

\section{Test System and Dynamic Models}

The test system and the dynamic models used in this work are well-established and publicly available. Consequently, this section will describe only the key characteristics and any modifications that were made.

\subsection{Test System} The IEEE Nordic Test System, detailed in \cite{pes_tr19, Implementation_Nordic}, is employed in this work due to its consideration of dynamics across various timescales. It includes typical fast-acting components like AVRs and PSSs, as well as slower-acting equipment such as OELs and LTCs. The system is publicly available in multiple formats on the IEEE PSDP website \cite{IEEE_PES_PSDP}.
Two modifications were made to the test system. First, the synchronous generators g6 and g7 in the central area were replaced by GFL PV plants, sized at 400~MVA and 200~MVA, respectively. These will be referred to as IBR\_6 and IBR\_7. Second, a 100~MVA GFM battery was considered at bus 1042, which is the same bus where IBR\_6 is connected, but only for one study case.

\subsection{GFL IBR Model}

For modeling GFL IBRs, we adopt the Western Electric Coordinating Council (WECC) REGC\_C model \cite{Converter_Model_V_Source, Pos_Seq_VSC}, which is widely available in several software tools. Key features of this model relevant to this work include the consideration of both PLL and inner current control dynamics. For more insights into the model's structure and its applicability to the phenomena discussed in this publication, refer to the author's previous works \cite{CDI_long_term, pabon2025grid}. In addition to the inner control, the outer control REEC\_D and plant control REPC\_A models were also considered. These models have been developed and validated by the members of the WECC Model Validation Subcommittee. Furthermore, we employ the singular perturbation  \cite{kokotovic1999singular} and modal analysis to examine the timescales in the GFL IBR model when connected to an infinite bus. The singular perturbation analysis reveals that the slowest states of the GFL system are primarily driven by the voltage controller state of REPC\_A and the PLL states of REGC\_C.

\subsection{GFM IBR Model}

The well-established WECC REGFM\_A1 model is used in this work to model GFM inverters. According to \cite{esig_gfm_modeling}, the model's development was a collaborative effort between the Pacific Northwest National Laboratory’s (PNNL) Directed Research and Development program and the Universal Interoperability for Grid-Forming Inverters (UNIFI) consortium, through collaboration between PNNL, Electric Power Research Institute (EPRI), and SMA Solar Technology. This positive-sequence model is capable of representing the characteristics of droop-controlled GFM IBRs and has been approved by the WECC Model Validation Subcommittee for use in interconnection-wide studies within the WECC. Details can be found in \cite{pacific2023model}.

\section{Numerical Studies and Main Results} 

One of the primary characteristics of the test system used in this work is the high power transfer from the North to the Central area. Under these conditions, the tripping of a transmission line, such as Line 4032-4044, can lead to long-term instability or collapse. This is a phenomenon that can be attributed mainly to slow dynamics, starting with the action of LTCs to control distribution voltages, which attempts to restore load consumption towards the pre-disturbance value which is unfeasible after the line trip. To control voltage, synchronous machines in the Central region increase their field current which eventually leads to the action of their OEL that limits the field current to protect the generator windings. These generators lose their voltage control capability. Depending on the considered scenario, the system reaches either instability, e.g., unacceptable low voltages, sustained oscillations, etc., or collapse, which is an abrupt fall of voltages resulting in a system blackout \cite{pes_tr19}. For this reason, this study investigates the tripping of said line as the primary event.

Four cases are presented in this work and are summarized in Table~\ref{table_cases}.
\begin{table}[!t]
\renewcommand{\arraystretch}{1.3}
\centering
\caption{Case Study Setups}
\label{table_cases}
\begin{tabular}{|l|c|c|c|c|}
\hline
\textbf{Case} & \textbf{Case 1} & \textbf{Case 2} & \textbf{Case 3} & \textbf{Case 4} \\
\hline
\textbf{GFL PV (IBRs 6 and 7)} & Yes & Yes & Yes & Yes \\
\hline
\textbf{PLL Bandwidth in Hz} & 8 & 8 & 4 & 8 \\
\hline
\textbf{Voltage Emergency Control} & No & Yes & Yes & Yes \\
\hline
\textbf{GFM Battery} & No & No & No & Yes \\
\hline
\end{tabular}
\end{table}
The ``Voltage Emergency Control" in Table~\ref{table_cases} corresponds to a typical countermeasure often implemented as conservation voltage reduction (CVR). This technique mitigates emergencies by reducing the voltage setpoint of LTCs to a predefined value, for example, 0.95 pu. This action takes advantage of the voltage sensitivity of residential loads to lower consumption during an emergency. While traditional methods use fixed voltage reductions, more advanced approaches exist, such as the one presented by the authors in \cite{pabon_emergency_control} and patented in Europe and the U.S. \cite{ospina2023electrical}. These methods offer a less intrusive scheme where the distribution voltage is not decreased by a fixed value. Activation of these emergency controls can be triggered by monitoring transmission voltage thresholds or, more effectively, by applying advanced methods like Local Identification of Voltage Emergency Situations (LIVES) or ``New LIVES" index \cite{NewLIVES}.


\subsection{Case 1 - GFL (Fast PLL) Without Voltage Emergency Control}

The evolution of transmission voltages after the contingency is presented in Fig.~\ref{fig:voltages_case_1}, which concludes that:

\begin{figure}[!b]
    \centering
    \includegraphics[width=1.0\linewidth]{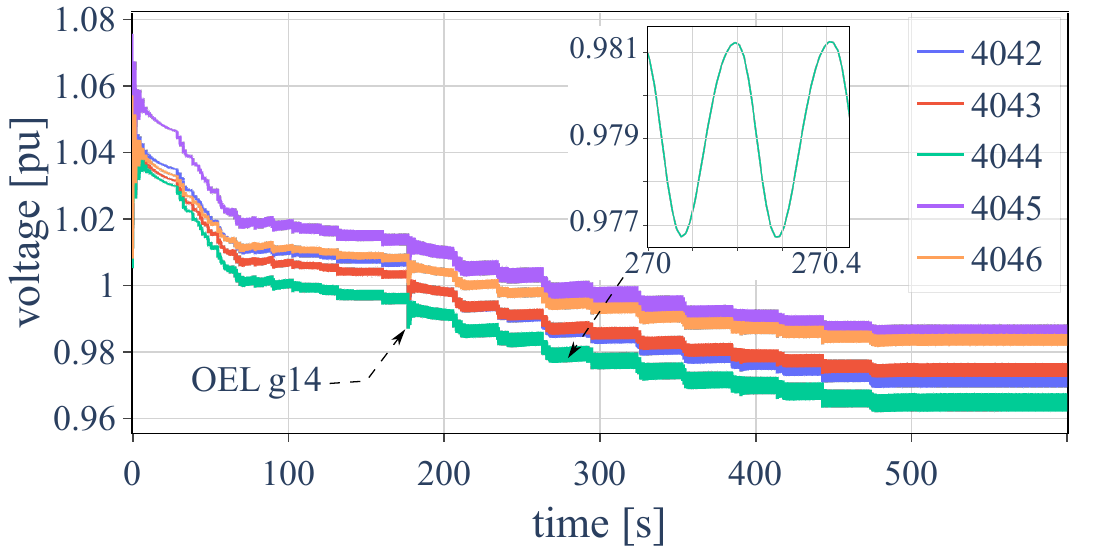}
    \caption{Evolution of voltage magnitudes in Case 1.}
    \label{fig:voltages_case_1}
\end{figure}

\begin{itemize}
\setlength\itemsep{1em}
    \item Despite the generating units' control efforts, the voltage cannot be restored to its pre-disturbance value.
    \item An undamped oscillation is already visible in the transmission voltages. This oscillation is more pronounced in the power output of IBRs, as will be shown shortly.
    \item The system ends up operating in a state where some synchronous generators have lost their voltage control capabilities. A critical event happens shortly before 200~s, when the OEL of generator g14 in the central area acts.
\end{itemize}



\begin{figure}[!t]
    \centering
    \includegraphics[width=0.9\linewidth]{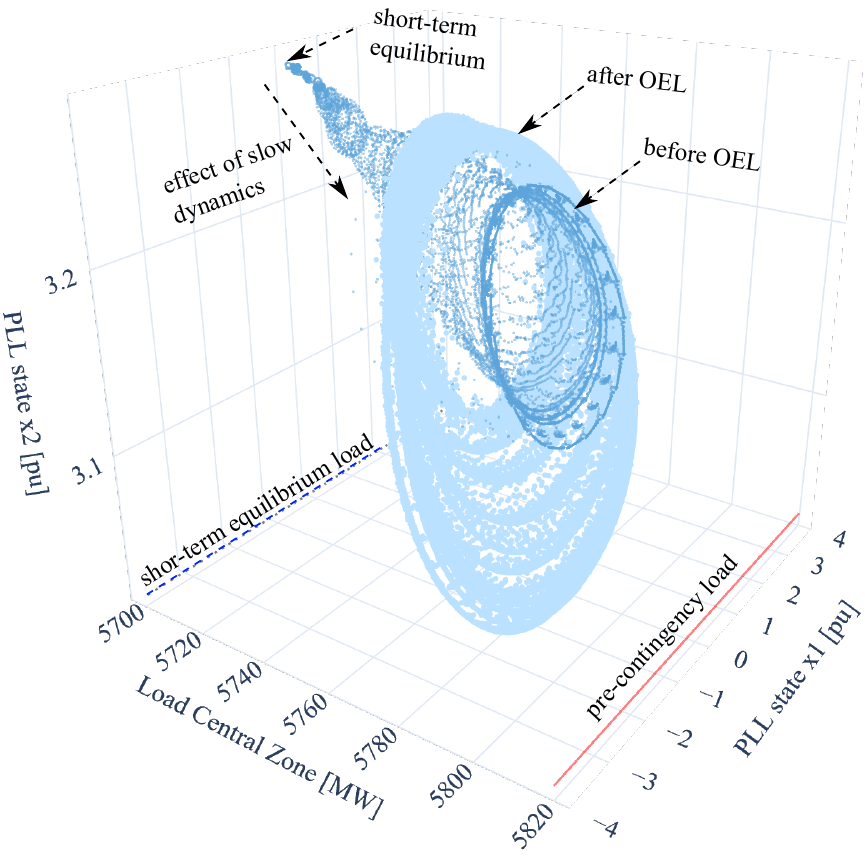}
    \caption{Phase-space trajectory in Case 1.}
    \label{fig:p_OEL_limit_cycle_case1}
\end{figure}

The observed undamped oscillation is not a direct consequence of the initial contingency; instead, it is induced by unstable slow dynamics (S-LT3). This is explained with Fig.~\ref{fig:p_OEL_limit_cycle_case1}, which shows the system response after short-term equilibrium. This figure presents substantial information and is explained step-by-step below.


Let's focus on the two axes where the PLL states of IBR\_6 are plotted. The state $x_1$ is associated with the proportional-integral controller, which is standard in most synchronous-reference frame PLLs (the block diagram can be found in \cite{CDI_long_term}). The state $x_2$ represents the estimated voltage phase. Plotting these two states provides a phase plane for the PLL dynamics. After the contingency, these states oscillate but quickly settle to a short-term equilibrium, which is marked in the figure. This would correspond to point \textbf{B} in Fig.~\ref{fig: manifolds_dont_intercept}.

Let us now focus on the ``Load Central Zone [MW]" axis. The original load consumption (before the contingency) is marked with a red line shortly before 5820~MW. When the system finds the aforementioned short-term equilibrium after the fault, the load consumption has been reduced to around 5700~MW (marked with a dashed line). The overall effect of slow dynamics led by LTC actions is to slowly restore the load from the dashed line towards the red line in Fig.~\ref{fig:p_OEL_limit_cycle_case1}. This is the equivalent of traveling from point \textbf{B} in Fig.~\ref{fig: manifolds_intercept} or Fig.~\ref{fig: manifolds_dont_intercept} towards the slow dynamics equilibrium manifold (exemplified with the orange line). In this case, the fast PLL dynamics lose stability as the system travels towards the slow dynamics equilibrium manifold (as in Fig.~\ref{fig: manifolds_dont_intercept}). The PLL dynamics encounter a \textit{Hopf} bifurcation that gives birth to a stable limit cycle on the PLL states phase plane. The oscillatory behavior becomes more pronounced the more the load consumption of the central zone is restored.
As previously mentioned, there is an interesting event shortly before 200~s. The OEL of synchronous machine g14 in the affected area operates. This inhibits the voltage control capability of the generator, further hindering the stability of the system. This is the reason why Fig.~\ref{fig:p_OEL_limit_cycle_case1} has two shades of blue. The evolution of the PLL states before the OEL of g14 acts is shown with dark blue. After the OEL action, the trace is shown with light blue. This is to show the significant effect that the OEL action has on the stability of faster PLL dynamics. Two things can be highlighted after the OEL action. First, the oscillatory behavior becomes even more evident with an even ``bigger" limit cycle. Second, the active power of the central zone is slightly reduced. This is due to the voltage dependency of loads after voltages in the central area are reduced due to the OEL action.
Figure \ref{fig:p_OEL_limit_cycle_case1} might be unconventional in the power system dynamics community, but it reveals the intrinsic interactions between slow and fast dynamics.

\begin{figure}[!b]
    \centering
    \begin{subfigure}[t]{0.49\linewidth} 
        \centering
        \includegraphics[width=1\linewidth]{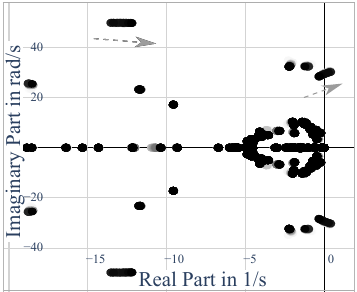}
        \caption{} 
        \label{fig:eigenvalues_participations:a}
    \end{subfigure}
    \hfill 
    \begin{subfigure}[t]{0.49\linewidth} 
        \centering
        \includegraphics[width=1\linewidth]{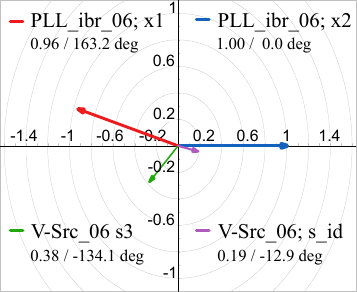}
        \caption{} 
        \label{fig:eigenvalues_participations:b}
    \end{subfigure}
    \caption{Eigenvalues (a) and participation factors (b).}
    \label{fig:eigenvalues_participations}
\end{figure}

\begin{figure}[!b]
    \centering
    \includegraphics[width=1\linewidth]{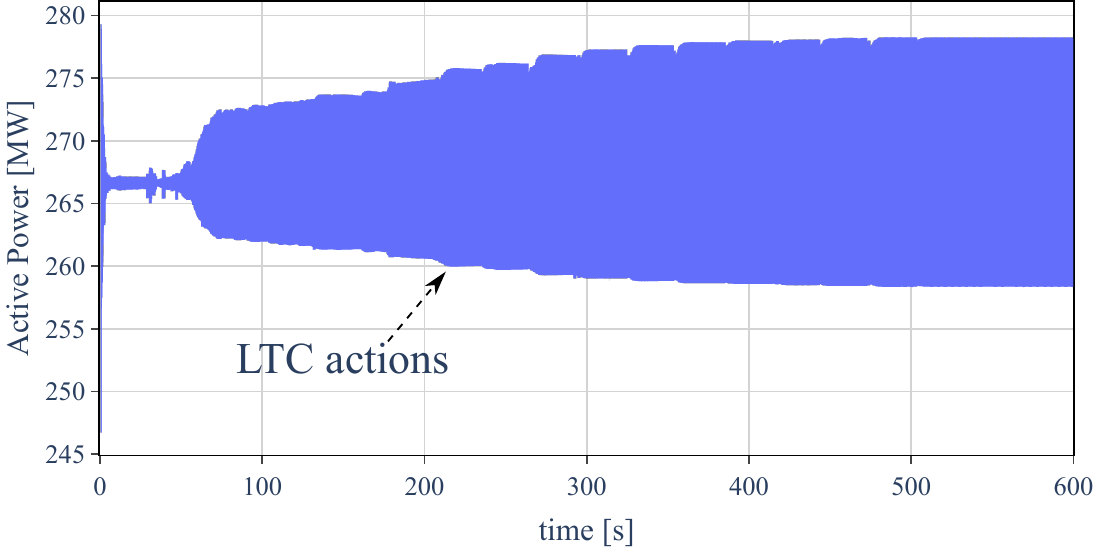}
    \caption{Active power of IBR\_6 in Case 1.}
    \label{fig:P_IBG_case1}
\end{figure}

\begin{figure}[!b]
    \centering
    \includegraphics[width=1\linewidth]{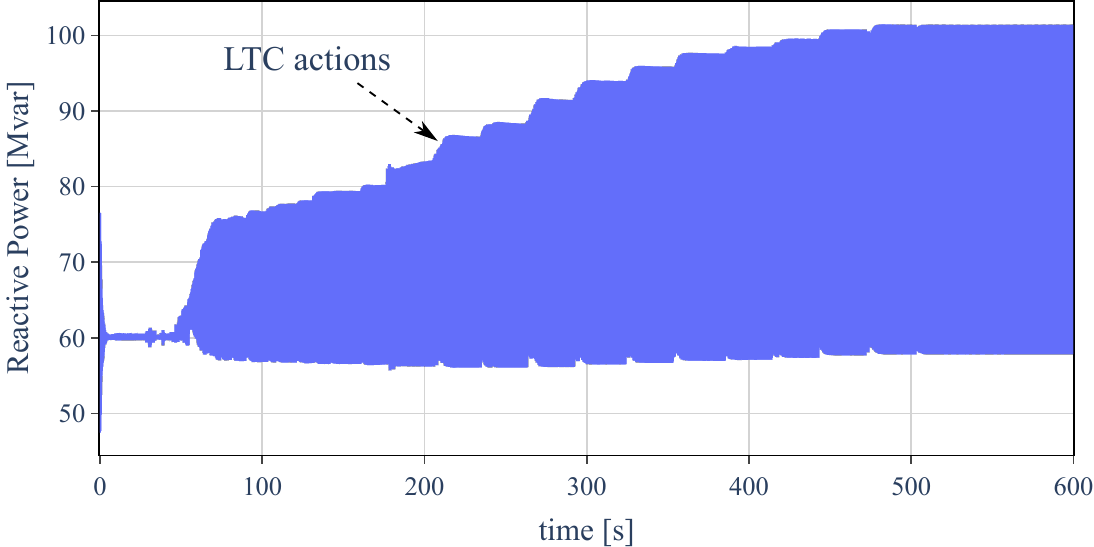}
    \caption{Reactive power of IBR\_6 in Case 1.}
    \label{fig:Q_IBG_case1}
\end{figure}

To corroborate that the PLL states are responsible for the S-LT3 resulting in a limit cycle, Fig.~\ref{fig:eigenvalues_participations} presents the eigenvalue trajectories and the participation factors for the undamped mode. In Fig.~\ref{fig:eigenvalues_participations:b}, only states with a contribution higher than 0.1 are shown. The PLL states exhibit the largest participation, while other states in the voltage source interface and inner current control show a smaller contribution. The S-LT3 instability leads to undamped oscillations in the IBR power output, which is visible in Figs.~\ref{fig:P_IBG_case1} and \ref{fig:Q_IBG_case1}.

\begin{figure}[!t]
    \centering
    \includegraphics[width=1\linewidth]{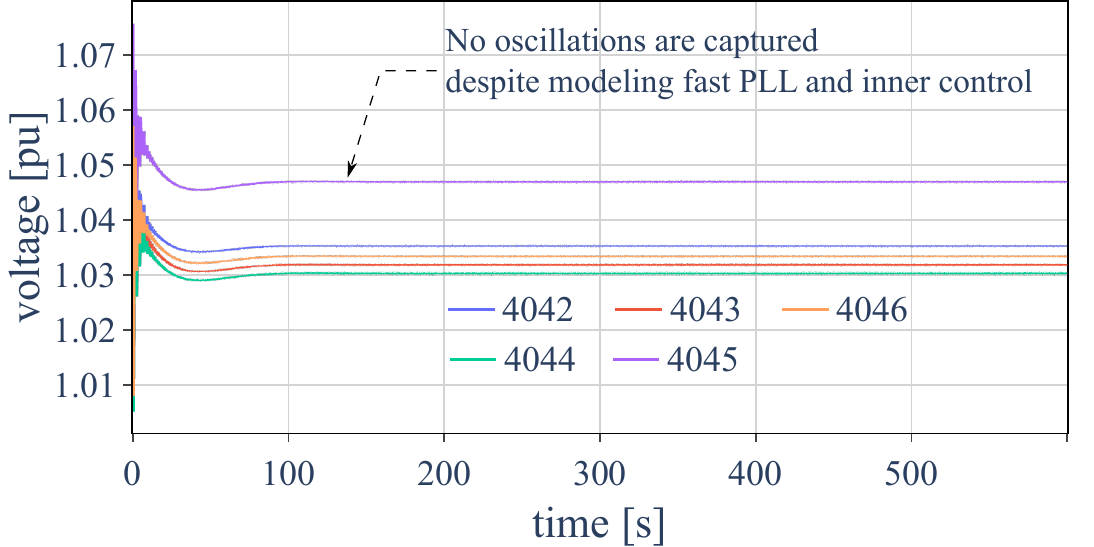}
    \caption{Results without considering multi-timescale dynamics.}
    \label{fig:voltages_no_LTC}
\end{figure}

Note that the S-LT3 instability captured in this section would be missed if multi-timescale dynamics are not considered. To corroborate this, Fig.~\ref{fig:voltages_no_LTC} presents the case in which the  PLL and the inner current control of IBRs are modeled. These are thought to be the key elements in capturing such forced oscillations. Nevertheless, the slow dynamic components (LTCs, OELs, etc.) will not be modeled.

The conclusion is clear: Even though the fast IBR dynamics were modeled, no oscillations were captured. This is because there are no mechanisms that make the system travel towards the slow-dynamics equilibrium manifold. In other words, the system will remain at point \textbf{B} of Fig.~\ref{fig: manifolds_dont_intercept}, giving a false sense of a stable equilibrium. Only when the slow and fast dynamics are modeled together do the interactions between both timescales become evident, and the S-LT3 instability is detected.


\subsection{Case 2 - GFL (Fast PLL) With Voltage Emergency Control}

\begin{figure}[!b]
    \centering
    \includegraphics[width=1\linewidth]{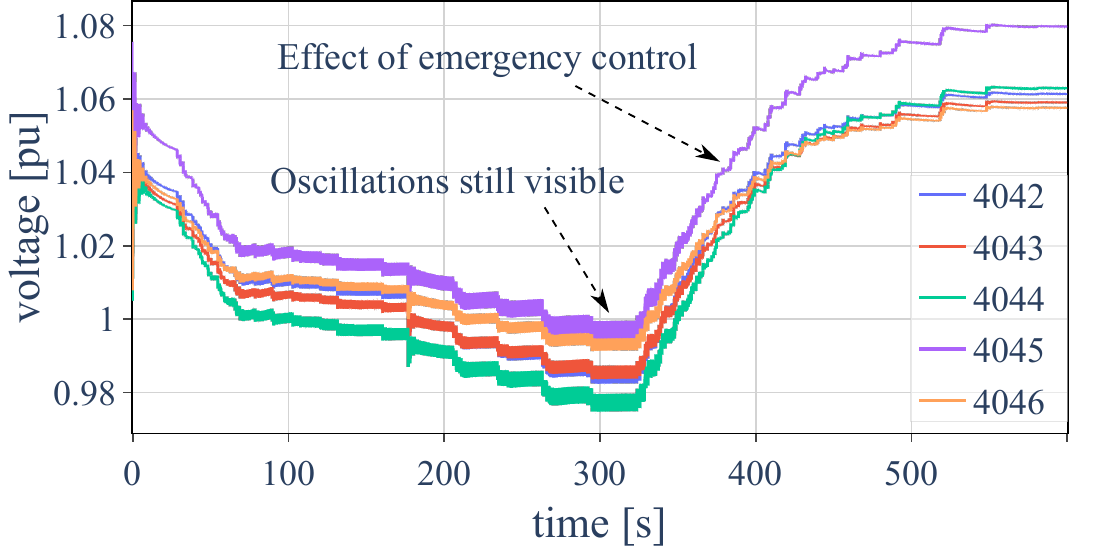}
    \caption{Evolution of voltage magnitudes in Case 2.}
    \label{fig:voltages_case2}
\end{figure}

The S-LT3 instability presented in the previous section has two issues that need to be corrected. First, the voltages should be recovered and the system stress needs to be relieved so that synchronous generators can regain their voltage control capability instead of being limited by their OEL. Second, the undamped oscillations need to be mitigated. Case 2 addresses the first issue. To do this, a classical CVR scheme is applied in which the voltage setpoint of LTCs is reduced by 5\% at $t=300 s$. The effect of such control can be observed in Fig.~\ref{fig:voltages_case2}.

\begin{figure}[!b]
    \centering
    \includegraphics[width=0.9\linewidth]{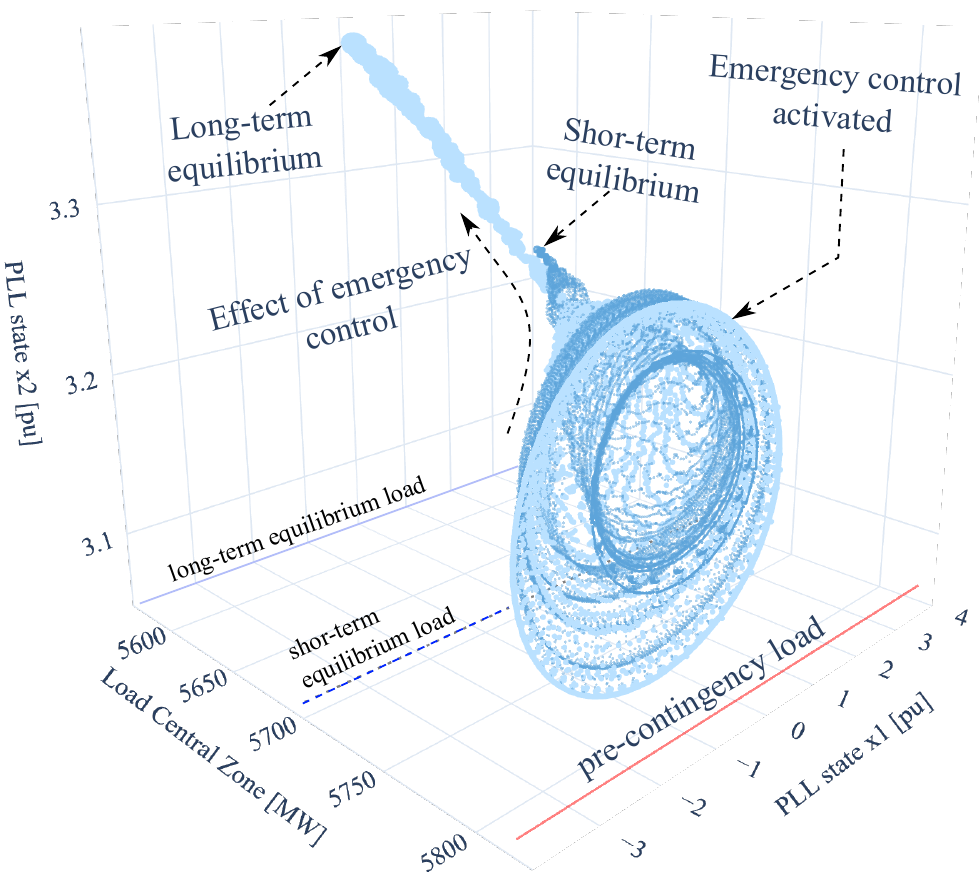}
    \caption{Phase-space trajectory after emergency control in Case 2.}
    \label{fig:p_limit_cycle_case2}
\end{figure}

Figure \ref{fig:p_limit_cycle_case2} presents the PLL phase plane together with the active power consumption in the Central zone. Note that the trace has two shades of blue. All results before the CVR scheme at $t=300 s$ are shown in dark blue. The slow dynamics again cause an increase in active power in the central area, from the dashed line (before 5700 MW) to the red line (after 5800 MW). The fast dynamics of the PLL oscillate exactly as shown in the previous section. The difference starts at $t\geq300 s$ when the CVR is activated. The results for this period are shown with a light blue trace. The CVR scheme reduces the power consumption as indicated by the direction arrow labeled ``Effect of emergency control" in Fig.~\ref{fig:p_limit_cycle_case2}. As the loading decreases, the voltage angle $x_2$ increases. The system ultimately reaches a long-term equilibrium, as indicated in the figure.

\begin{figure}[!b]
    \centering
    \includegraphics[width=1\linewidth]{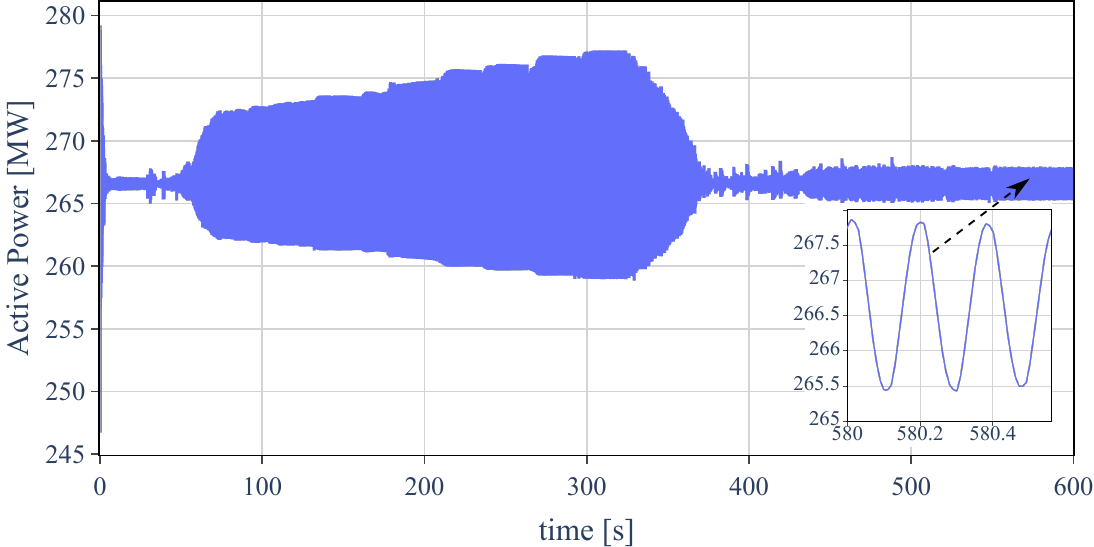}
    \caption{Active power of IBR\_6 in Case 2.}
    \label{fig:P_IBG_case2}
\end{figure}

Nevertheless, this is an unstable equilibrium in which oscillations still exist (recall that Case 2 is supposed to fix only the first of the two mentioned issues). The existence of oscillations at the long-term equilibrium is difficult to see in Fig.~\ref{fig:p_limit_cycle_case2} due to the scaling of the figure. A better way to see that oscillations still exist is presented in Fig.~\ref{fig:limit_cycles_case2_vs_case3} (blue trace). This is the limit cycle on the PLL phase plane at the long-term equilibrium. In this case, a one-second trajectory on the phase plane is plotted, clearly showing the existence of a limit cycle. The oscillatory behavior is clearly visible on the IBR power output, as shown in Fig.~\ref{fig:P_IBG_case2}.

The following sections explore two ways to mitigate the oscillatory behavior of fast dynamics induced by unstable slow dynamics.


\subsection{Case 3 - GFL (Slow PLL) With Voltage Emergency Control}

\begin{figure}[!b]
    \centering
    \includegraphics[width=1.0\linewidth]{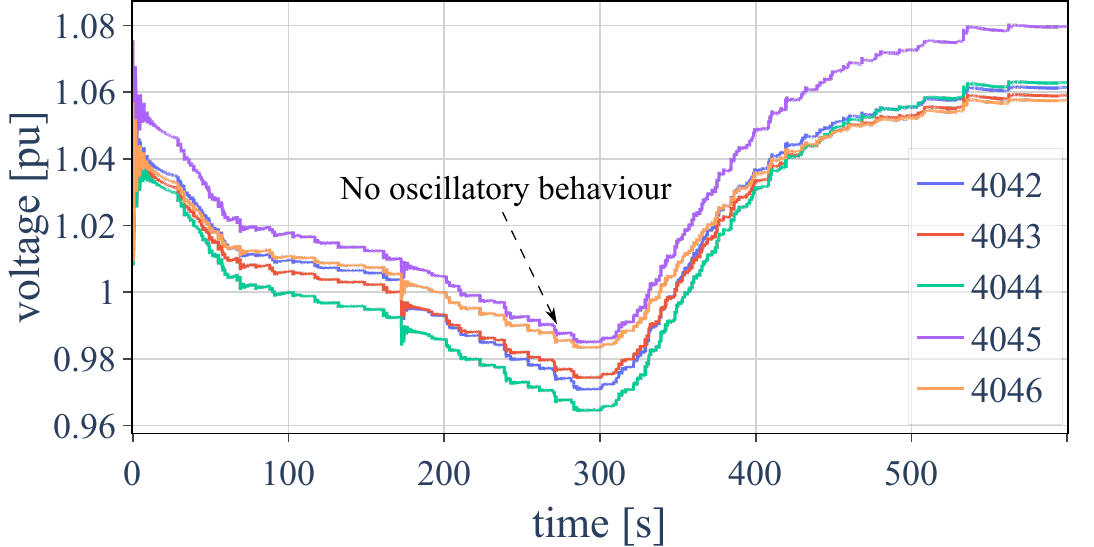}
    \caption{Evolution of voltage magnitudes in Case 3.}
    \label{fig:voltages_case3}
\end{figure}

The first method is to reduce the bandwidth of the PLL from 8~Hz to 4~Hz. This is a well-known strategy to reduce the risk of forced oscillations, although it has its limitations. The evolution of transmission voltages is presented in Fig.~\ref{fig:voltages_case3}, in which the oscillatory behavior is no longer perceptible.

\begin{figure}[!t]
    \centering
    \includegraphics[width=0.6\linewidth]{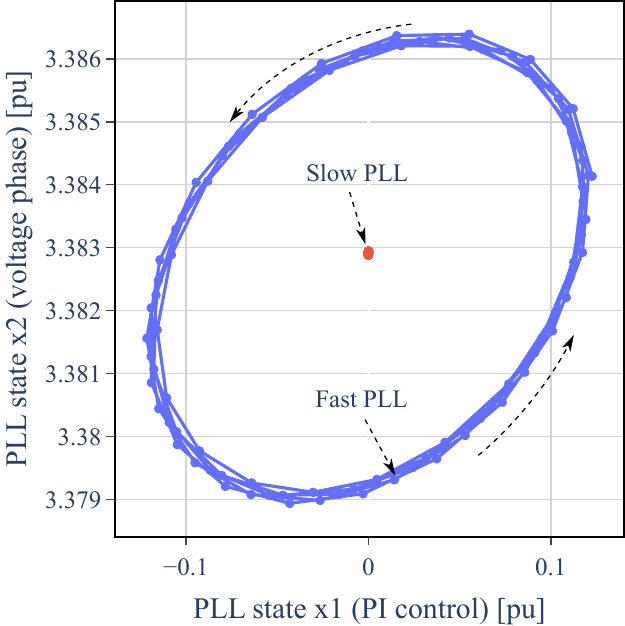}
    \caption{Limit cycles ($t=[599, 600]$) with fast PLL in Case 2 compared to stable
long-term equilibrium with slow PLL in Case 3.}
    \label{fig:limit_cycles_case2_vs_case3}
\end{figure}

To corroborate the non-existence of oscillations, the trajectory on the phase plane in Fig.~\ref{fig:limit_cycles_case2_vs_case3} is now superimposed with the results of the case with the 4~Hz-bandwidth PLL (marked as ``Slow PLL"). This confirms that the limit cycle disappears and the system reaches a stable long-term equilibrium.

Note that although decreasing the PLL bandwidth improves the dynamic performance in this case, it will slow down the overall control response of the IBR. This requires a compromise between avoiding oscillatory behavior and controller response time so that it meets interconnection requirements, such as rise times.
An alternative, more costly, solution is presented in the following section.


\subsection{Case 4 - GFL (Fast PLL) Plus GFM Battery}

The IBR\_6 represents a photovoltaic generator. While it is tempting to replace its control architecture with GFM control, this may not be a realistic assumption. Most state-of-the-art commercial GFM inverters need some form of energy storage, such as a battery or an HVDC link. Therefore, instead of replacing the existing PV GFL IBR\_6, this section considers a 100~MVA GFM battery connected to bus 1042, the same location where the substation transformer of IBR\_6 is connected. This section is not a direct comparison between GFL and GFM but intends to provide insights into how GFM technology can enhance dynamic performance while still allowing GFL IBRs to operate stably even with a relatively high PLL bandwidth.

\begin{figure}[!b]
    \centering
    \includegraphics[width=1.0\linewidth]{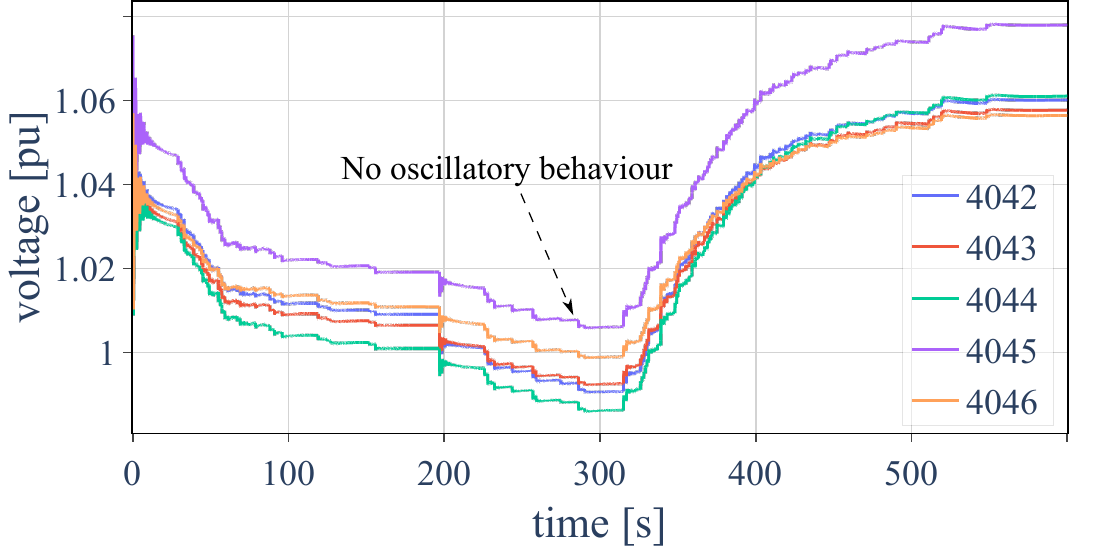}
    \caption{Evolution of voltage magnitudes in Case 4.}
    \label{fig:V_sim_GFM_fig}
\end{figure}

The voltage evolution is presented in Fig.~\ref{fig:V_sim_GFM_fig}. By monitoring only transmission voltages, the dynamic performance is very similar to that in Case 3, which simply considered a lower bandwidth PLL without the need for a new GFM battery (a comparison between Figs.~\ref{fig:voltages_case3} and \ref{fig:V_sim_GFM_fig} confirms this point). Nevertheless, the solution proposed in Case 3 is known to have limitations in extreme cases of low short circuit ratio (SCR), e.g., $SCR<1.2$. Case 4 has higher resilience since it considers a new asset that makes it possible to supply 100~MW locally without the need for power transfer through long distances in an N-1 system.

\begin{figure}[!t]
    \centering
    \includegraphics[width=1.0\linewidth]{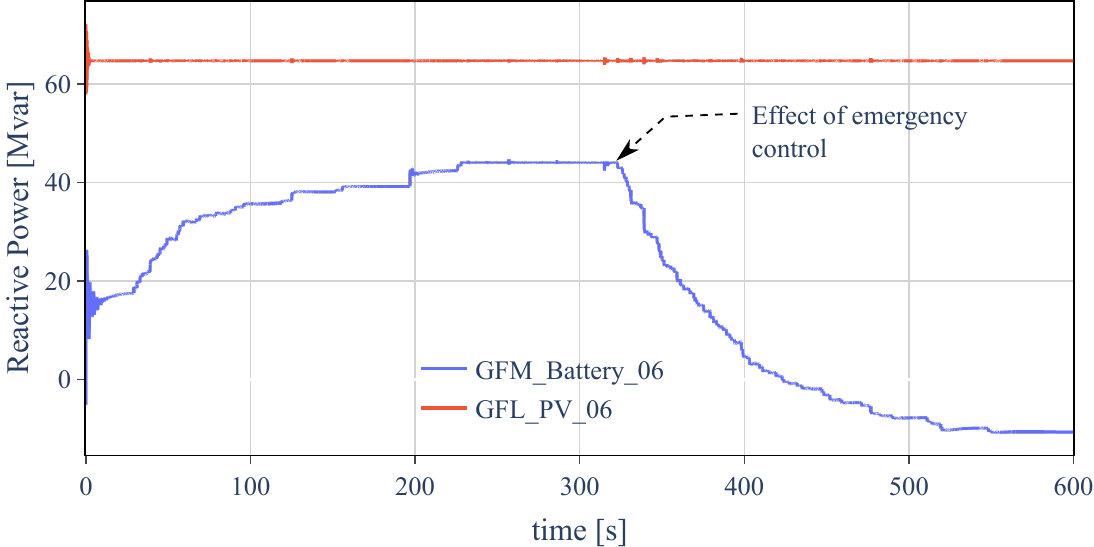}
    \caption{Reactive power of GFL and GFM in Case 4.}
    \label{fig:Q_IBGs_sim_GFM_fig}
\end{figure}

The reactive power of the GFM battery and GFL PV are shown in Fig.~\ref{fig:Q_IBGs_sim_GFM_fig}. It shows that the GFM control overtakes the voltage regulation, with the battery quickly supplying the necessary reactive power support while the GFL IBR keeps its reactive power infeed constant. Once the CVR scheme is activated, the load consumption in the affected area decreases, and the GFM battery can reduce its reactive power. In fact, at the end of the emergency control, the battery consumes reactive power to avoid overvoltages, which could be a potential issue with aggressive CVR.

\begin{figure}[!h]
    \centering
    \includegraphics[width=0.9\linewidth]{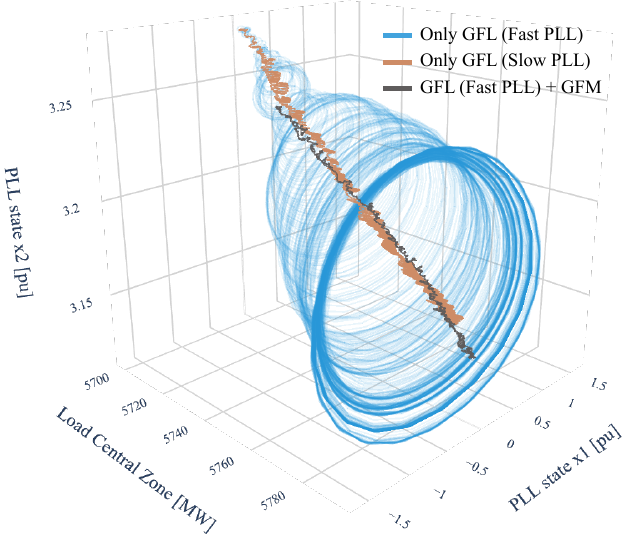}
    \caption{Phase-space trajectory with and without GFM battery.}
    \label{fig:p_limit_cycle_case1_vs_GFM_vs_case4}
\end{figure}

The phase plane trajectory with and without a GFM battery is shown in Fig.~\ref{fig:p_limit_cycle_case1_vs_GFM_vs_case4}. For legibility reasons, the trajectory is shown only until the critical point shortly before the OEL of g14 acts. After this point, the CVR will have an effect similar to the one in Fig.~\ref{fig:p_limit_cycle_case2}. The main goal of Fig.~\ref{fig:p_limit_cycle_case1_vs_GFM_vs_case4} is to confirm that both approaches in Cases 3 and 4 avoid the appearance of limit cycles as the long-term dynamics evolve. Case 1 is also shown as a reference.

\balance

\section{Conclusions and Future Work}

Power networks have dynamics that evolve at different timescales and have intrinsic interactions. The dynamics of one timescale can induce instabilities in those of a different timescale. This work highlights the importance of multi-timescale simulation and exemplifies a case of S-LT3 instability. It shows that the oscillatory behavior of fast IBR dynamics is induced by slower dynamics such as LTCs and OELs. The paper demonstrates that if multi-timescale simulation is not considered, this instability would be overlooked, and the system could be incorrectly deemed stable. This work also exemplifies how GFM technology can complement GFL controls by enabling stable operation of GFL IBRs without the need to slow down their control response.

This work exclusively used positive-sequence time-domain simulation to confirm that the interaction between slow and fast dynamics is significant even within the timescales that can be captured by this simulation method. While a wider range of dynamics can be captured by EMT simulations, ongoing work by the authors utilizes RMS-EMT co-simulation to further investigate crucial interdependencies between dynamics of different timescales.




\end{document}